\documentclass{article}

\def\.#1{\dot #1}

\def\C{\mathcal{C}}

\def\E{\mathcal{E}}

\def\L{\mathcal{L}}

\def\R{{\bf R}}  

\def\sse{\subseteq}
\def\ss{\subset}
\def\pa{\partial}

\def\=#1{{\widetilde #1}}
\def\.#1{\dot #1}
\def\^#1{\widehat #1}

\def \wt#1{{\widetilde #1}}
\def\sse{\subseteq}

\def\d{{\rm d}}       

\def\({\left(}
\def\){\right)}
\def\[{\left[}
\def\]{\right]}

\def\a{\alpha}

\def\ga{\gamma}
\def\de{\delta}   

\def\phi{\varphi}
\def\la{\lambda}
\def\La{\Lambda}

\def\vth{\vartheta}

\def\phi{\varphi}
\def\Ga{\Gamma}
\def\De{\Delta}

\def\w{\wedge}

\def\interno{\hskip 2pt \vbox{\hbox{\vbox to .18
truecm{\vfill\hbox to .25 truecm
{\hfill\hfill}\vfill}\vrule}\hrule}\hskip 2 pt}

\begin{document}

\title{\bf Lambda and mu-symmetries}

\author{Giuseppe Gaeta \\
{\it Dipartimento di Matematica, Universit\`a di Milano} \\
{\it via Saldini 50, I--20133 Milano (Italy)} \\
{\tt gaeta@mat.unimi.it}}

\date{{\it Contributed paper to SPT2004 Proceedings}}

\maketitle

\noindent {\bf Summary.} Lambda-symmetries of ODEs were discussed
by C. Muriel in her talk at SPT2001. Here we provide a geometrical
characterization of $\la$-pro\-long\-a\-ti\-ons, and a
generalization of these -- and of $\la$-symmetries -- to PDEs and
systems thereof.

\section*{Introduction}

Symmetry analysis is a standard and powerful method in the
analysis of differential equations, and in the determination of
explicit solutions of nonlinear ones.

It was remarked by Muriel and Romero \cite{MR} (see also the work
by Pucci and Saccomandi \cite{PuS}) that for ODEs the notion of
symmetry can be somehow relaxed to that of {\it lambda-symmetry}
(see below), still retaining the relevant properties for symmetry
reduction and hence for the construction of explicit solutions.
Their work was presented at SPT2001 \cite{Murspt}, raising
substantial interest among participants.

Here I report on some recent work \cite{CGM,GM,GML} which sheds
some light on ``lambda-symmetries'', and extends them to PDEs --
and systems thereof -- as well; as the central objects here are
not so much the functions $\la$, but some associated one-forms
$\mu$, these are called ``mu-symmetries''.

The work reported here was conducted together with Giampaolo Cicogna and Paola Morando; I would like to thank them, as well as other friends (J.F. Cari\~nena, G. Marmo, M.A. Rodr\'{\i}guez) with whom I discussed these topics in the near past. It is also a pleasure to thank C. Muriel and G. Saccomandi for privately communicating their work on $\la$-symmetries and raising my interest in the topic.

\section{Standard prolongations}

Let us consider equations with $p$ independent variables $(x^1,...,x^p) \in B = \R^p$ and $q$ dependent ones, $(u^1,...,u^q) \in F = \R^q$. The corresponding phase space will be $M = B \times F$; more precisely, this is a trivial bundle $(M,\pi,B)$.

With the notation $\pa_i := \pa / \pa x^i$ and $\pa_a := \pa / \pa u^a$, a Lie-point vector field in $M$ will be written as
$$ X \ = \ \xi^i (x,u) \, \pa_i \ + \ \phi^a (x,u ) \, \pa_a \ .  \eqno(1) $$
We also write, with $J$ a multiindex of lentgth $|J| = j_1 + ... + j_q$, $\pa_a^J := \pa / \pa u^a_J$. Then a vector field in the $n$-th jet bundle $J^n M$ will be written (sum over $J$ being limited to $0 \le |J| \le n$) as
$$ Y \ = \ \xi^i \, \pa_i \ + \ \Psi^a_J \, \pa_a^J \ . \eqno(2) $$

The jet space $J^n M$ is equipped with a {\bf contact structure}, described by the {\bf contact forms}
$$ \vth^a_J \ := \ \d u^a_J \ - \ u^a_{J,i} \, \d x^i \ \ \ \ \ (|J| \le n-1 ) \ . \eqno(3) $$

Denote by $\E$ the $\C^\infty (J^n M) $ module generated by these $\vth^a_J$. Then we say that $Y$ preserves the contact structure if and only if, for all $\vth \in \E$,
$$ \L_Y (\vth ) \ \in \ \E \ . \eqno(4) $$
As well known, this is equivalent to the requirement that the coefficients in (2) satisfy the (standard) {\bf prolongation formula}
$$ \Psi^a_{J,i} \ = \ D_i \, \Psi^a_J \ - \ u^a_{J,m} \, ( D_i \, \xi^m ) \ . \eqno(5) $$

We note, for later reference, that for scalar ODEs formula (5) is rewritten more simply, with obvious notation, as
$$ \Psi_{k+1} \ = \ D_x \, \Psi_k \ - \ u_{k+1} \, ( D_x \, \xi ) \ . \eqno(6) $$

We also recall that the vector field $Y$ is the prolongation of
$X$ if $Y$ satisfies (4) and coincides with $X$ when restricted to
$M$; $X$ is a symmetry of a differential equation (or system of
differential equations) $\De$ of order $n$ in $M$ if its $n$-th
prolongation $Y$ is tangent to the solution manifold $S_\De \ss
J^n M$, see standard references on the subject
\cite{CGspri,Gae,Kam,Kra,Olv,Ste,Win}.

Note that condition (4) is also equivalent to conditions involving
the commutator of $Y$ with the total derivative operators $D_i$;
in particular, it is equivalent to either one of
$$ [D_i , Y ] \interno \vth \ = \ 0 \ \ \ \forall \vth \in \E \ ; \eqno(7') $$
$$ [D_i , Y ] \ = \ h_i^m \, D_m \ + V \ , \eqno(7'') $$
with $h_i^m \in \C^\infty (J^n M)$ and $V$ a vertical vector field in $J^n M$ seen as a bundle over $J^{n-1} M$.

\section{Lambda-prolongations}

\subsection{The work of Muriel and Romero}

In 2001, C. Muriel and J.L. Romero \cite{MR}, analyzing the case
where $\De$ is a scalar ODE,  noticed a rather puzzling fact.

They substitute the standard prolongation formula (6) with a
``lambda-prolongation'' formula
$$ \Psi_{k+1} \ = \ (D_x + \la ) \, \Psi_k \ - \ u_{k+1} \, ( D_x + \la ) \, \xi  \ ; \eqno(8) $$
here $\la$ is a real $C^\infty$ function defined on $J^1 M$ (or on
$J^k M$ if one is ready to deal with generalized vector fields).
Let us now agree to say that $X$ is a ``lambda-symmetry'' of $\De$
if its ``lambda-prolongation'' $Y$ is tangent to the solution
manifold $S_\De \ss J^n M$.

Then, it turns out that ``lambda-symmetries'' are as good as
standard symmetries for what concerns symmetry reduction of the
differential equation $\De$ and hence determination of its
explicit solutions. As pointed out by Muriel and Romero, it is
quite possible to have equations which have no standard
symmetries, but possess lambda-symmetries and can therefore be
integrated by means of their approach; see their works
\cite{MR,MR2} for examples.

\subsection{The work of Pucci and Saccomandi}

In 2002, Pucci and Saccomandi \cite{PuS} devoted further study to
lambda-symmetries, and stressed a very interesting geometrical
property of lambda-prolongations: that is, lambda-prolonged vector
fields in $J^n M$ can be characterized as the {\it only} vector
fields  in $J^n M$ which have the same characteristics as some
standardly-prolonged vector field.

We stress that $Y$ is the lambda-prolongation of a vector field $X$ in $M$, then the characteristics of $Y$ will not be the same as those of the standard prolongation $X^{(n)}$ of $X$, but as those of the standard prolongation $\wt{X}^{(n)}$ of a different (for $\la$ nontrivial) vector field $\wt{X}$ in $M$.

This property can also be understood by recalling (4) and making use of a general property of Lie derivatives: indeed, for $\a$ any form on $J^n M$,
$$ \L_{\la Y} (\a) \ = \ \la Y \interno \d \a \, + \, \d (\la Y \interno \a ) \ = \ \la \, \L_Y (\a ) \, + \, \d \la \w (Y \interno \a ) \ . \eqno(9) $$

\subsection{The work of Morando}

It was noted \cite{GM,MR} that lambda-prolongations can be given a
characterization similar to the one discussed in remark 1 for
standard prolongations; that is, with $h_i^m$ and $V$ as above,
(8) is equivalent to either one of $ [D_x , Y ] \interno \vth =
\la (Y \interno \vth )$ for all $\vth \in \E$, and $ [D_x , Y ] =
\la  Y  +  h_i^m D_m + V $.

This, as remarked by Morando, also allows to provide a
characterization of lambda-prolonged vector fields in terms of
their action on the contact forms, analogously to (4). In this
context, it is natural to focus on the one-form $ \mu := \la \d x
$; note this is horizontal for $J^n M$ seen as a bundle over $B$,
and obviously satisfies $D \mu = 0$, with $D$ the total exterior
derivative operator. Then, {\it $Y$ is a lambda-prolonged vector
field if and only if $ \L_Y (\vth ) + (Y \interno \vth ) \mu \in
\E $ for all $\vth \in \E$}.

\section{Mu-prolongations; mu-symmetries for PDEs}

The result given above immediately opens the way to extend
lambda-symmetries to PDEs \cite{GM}. As here the main object will
be the one-form $\mu$, we prefer to speak of ``mu-prolongations''
and ``mu-symmetries''. Let
$$ \mu \ := \ \la_i \, \d x^i \eqno(10) $$ be a semibasic one-form on $J^n M, \pi_n , B)$, satisfying $D \mu = 0$. Then we say that the vector field $Y$ in $J^n M$ $\mu$-preserves the contact structure if and only if, for all $\vth \in \E$,
$$ \L_Y (\vth ) \ + \ (Y \interno \vth ) \, \mu \ \in \ \E \ . \eqno(11) $$
Note that $D \mu = 0$ means $D_i \la_j = D_j \la_i $ for all $i,j$; hence {\it locally} $\mu = D \Phi$ for some smooth real function $\Phi$.

With standard computations \cite{GM}, one obtains that (11)
implies the {\bf scalar $\mu$-prolongation formula}
$$ \Psi_{J,i} \ = \ (D_i + \la_i ) \, \Psi_J \ - \ u_{J,m} \, ( D_i + \la_i ) \, \xi^m  \ . \eqno(12) $$

Let $Y$ as in (2) be the $\mu$-prolongation of the Lie-point
vector field $X$ (1), and write the standard prolongation of the
latter as $ X^{(n)} = \xi^i  \pa_i + \Phi_J  \pa_u^J$; note that
$\Psi_0 = \Phi_0 = \phi$. We can obviously always write $\Psi_J =
\Phi_J + F_J$, and $F_0 = 0$. Then it can be proved \cite{GM} that
the difference terms $F_J$ satisfy the recursion relation
$$ F_{J,i} \ = \ (D_i + \la_i ) F_j + \la_i D_J Q \eqno(13) $$
where $Q := \phi - u_i \xi^i$ is the characteristic
\cite{Gae,Olv,Ste} of the vector field $X$.

This shows at once that {\it the $\mu$-prolongation of $X$
coincides with its standard prolongation on the $X$-invariant
space $I_X$}; indeed, $I_X \ss J^n M$ is the subspace identified
by $D_J Q = 0$ for all $J$ of length $0 \le |J| < n$. It follows
that the standard PDE symmetry reduction method \cite{Gae,Olv,Ste}
works equally well when $X$ is a $\mu$-symmetry of $\De$ as in the
case where $X$ is a standard symmetry of $\De$; see our work
\cite{GM} for examples.

The concept of $\mu$-symmetries is also generalized to an analogue
of standard conditional and partial symmetries \cite{CGpar,CiK},
i.e. partial (conditional) $\mu$-symmetries \cite{CGM}.

\section{Mu-symmetries for systems of PDEs}

The developements described in the previous section do not include
the case of (systems of) PDEs for several dependent variables,
i.e. the case with $q>1$ in our present notation. This was dealt
with in a recent work \cite{GM}, to which we refer for details.

To deal with this case, it is convenient to see the contact forms
$\vth^a_J$, see (3), as the components of a vector-valued contact
form \cite{Str} $\vth_J$. We will denote by $\Theta$ the module
over $q$-dimensional smooth matrix functions generated by the
$\vth_J$, i.e. the set of vefctor-valued forms which can be
written as $\eta = (R_J)^a_b \vth^b_J $ with $R_J : J^n M \to
Mat(q)$ smooth matrix functions.

Correspondingly, the fundamental form $\mu$ will be a horizontal
one-form with values in the Lie algebra $g \ell(q)$ (the algebra
of the group $GL(q)$, consisting of non-singular $q$-dimensional
real matrices) \cite{Str}. We will thus write
$$ \mu \ = \ \La_i \, \d x^i \eqno(14) $$
where $\La_i$ are smooth matrix functions satisfying additional compatibility conditions discussed below.

We will say that the vector field $Y$ in $J^n M$ $\mu$-preserves the vector contact structure $\Theta$ if, for all $\vth \in \Theta$,
$$ \L_Y (\vth ) \ + \ \( Y \interno (\La_i)^a_b \vth^b \) \ \d x^i  \ \in \ \Theta \ . \eqno(15) $$

In terms of the coefficients of $Y$, see (2), this is equivalent to the requirement that the $\Psi^a_J$ obey the {\bf vector $\mu$-prolongation formula}
$$ \Psi^a_{J,i} \ = \ (\nabla_i)^a_b \, \Psi^b_J \ - \ u^b_{J,m} \, [( \nabla_i)^a_b \, \xi^m ] \ , \eqno(16) $$
where we have introduced the (matrix) differential operators
$$ \nabla_i \ := \ I \, D_i \ + \ \La_i \ . \eqno(17) $$

If again we consider a vector field $Y$ as in (2) which is the $\mu$-prolongation of a Lie-point vector field $X$, and write the standard prolongation of the latter as $ X^{(n)} = \xi^i  \pa_i + \Phi^a_J  \pa_a^J$ (with $\Psi^a_0 = \Phi^a_0 = \phi^a$), we can write $\Psi^a_J = \Phi^a_J + F^a_J$, with $F^a_0 = 0$. Then the difference terms $F_J$ satisfy the recursion relation
$$ F^a_{J,i} \ = \ \de^a_b \[ D_i (\Ga^J)^b_c \] (D_j Q^c ) \, + \, (\La_i)^a_b \[ (\Ga^J)^b_c (D_J Q^c) + D_J Q^b \] \, \eqno(18) $$
where $Q^a := \phi^a - u^a_i \xi^i$, and $\Ga^J$ are certain
matrices (see ref. \cite{GM} for the explicit expression). This,
as for the scalar case, shows that the $\mu$-prolongation of $X$
coincides with its standard prolongation on the $X$-invariant
space $I_X$; hence, again, the standard PDE symmetry reduction
method works equally well for $\mu$-symmetries (defined in the
obvious way) as for standard ones. See ref. \cite{GM} for
examples.

\section{Compatibility condition, and gauge \\ equivalence}

As mentioned above the form $\mu$, see (14), is not arbitrary: it must satisfy a compatibility condition (this guarantees the $\Psi^a_J$ defined by (16) are uniquely determined), expressed by
$$ \[ \nabla_i , \nabla_k \] \ \equiv \ D_i \La_k \, - \, D_k \La_i \ + \ [\La_i , \La_k ] \ = \ 0 \ . \eqno(19) $$

It is quite interesting to remark \cite{CGM} that this is nothing
but the coordinate expression for the horizontal Maurer-Cartan
equation
$$ D \mu \ + \ {1 \over 2} \ [ \mu , \mu ] \ = \ 0 \ . \eqno(20) $$

Based on this condition, and on classical results of differential
geometry \cite{Sha}, it follows that locally in any contractible
neighbourhood $A \sse J^n M$, there exists $\ga_A : A \to GL(q)$
such that (locally in $A$) $\mu$ is the Darboux derivative of
$\ga_A$.

In other words, any $\mu$-prolonged vector field is {\it locally}
gauge-equivalent to a standard prolonged vector field \cite{CGM},
the gauge group being $GL(q)$.

It should be mentioned that when $J^n M$ is topologically
nontrivial, or $\mu$ present singular points, one can have
nontrivial $\mu$-symmetries; this is shown by means of very
concrete examples in our  recent work \cite{CGM}.

Note that when we consider symmetries of a given equation $\De$,
the compatibility condition (20) needs to be satisfied only on
$S_\De \sse J^n M$. When indeed $\mu$ is not satisfying everywhere
(20), $\mu$-symmetries can happen to be gauge-equivalent to
standard {\it nonlocal symmetries} of exponential form; see again
ref. \cite{CGM} for details.


\begin{thebibliography}{99}

\bibitem{CiK} G. Cicogna, ``A discussion on the different notions of
symmetry of differential equations'', {\it Proc. Inst. Math.
N.A.S. Ukr.} {\bf 50} (2004), 77-84; ``Weak symmetries and
symmetry adapted coordinates in differential problems'', {\it Int.
J. Geom. Meth. Mod. Phys.} {\bf 1} (2004), 23-31

\bibitem{CGspri} G. Cicogna and G. Gaeta, {\it Symmetry and perturbation theory in nonlinear dynamics}, Springer 1999

\bibitem{CGpar} G. Cicogna and G. Gaeta, ``Partial Lie-point symmetries of differential equations'', {\it J. Phys. A} {\bf 34} (2001), 491-512

\bibitem{CGM} G. Cicogna, G. Gaeta and P. Morando, ``On the relation between standard and $\mu$-symmetries for PDEs'', preprint 2004

\bibitem{Gae} G. Gaeta, {\it Nonlinear symmetries and nonlinear equations}, Kluwer 1994

\bibitem{GM} G. Gaeta and P. Morando, ``On the geometry of lambda-symmetries and PDEs reduction'', {\it J. Phys. A} {\bf 37} (2004), 6955-6975

\bibitem{GML} G. Gaeta and P. Morando, ``PDEs reduction and $\la$-symmetries'', to appear in {\it Note di Matematica}

\bibitem{Kam} N. Kamran, ``Selected topics in the geometrical study of differential equations'', A.M.S. 2002

\bibitem{Kra} I.S. Krasil'schik and A.M. Vinogradov eds., {\it Symmetries and conservation laws for differential equations of mathematical physics}, A.M.S. 1999

\bibitem{MR} C. Muriel and J.L. Romero, ``New method of reduction for ordinary differential equations'', {\it IMA Journal of Applied mathematics} {\bf 66} (2001), 111-125

\bibitem{Murspt} C. Muriel and J.L. Romero, ``$C^\infty$ symmetries and equations with
symmetry algebra $SL(2,R)$'', in: {\it Symmetry and Perturbation
Theory (SPT2001)}, D. Bambusi, M. Cadoni and G. Gaeta eds., World
Scientific 2001

\bibitem{MR2} C. Muriel and J.L. Romero, ``$C^\infty$ symmetries and
nonsolvable symmetry algebras'', {\it IMA Journal of Applied
mathematics} {\bf 66} (2001), 477-498; `` Integrability of
equations admitting the nonsolvable symmetry algebra $so(3,r)$'',
{\it Studies in Applied Mathematics} {\bf 109} (2002), 337-352;
``$C^\infty$ symmetries and reduction of equations without
Lie-point symmetries'', {\it Journal of Lie theory} {\bf 13}
(2003), 167-188; M.L. Gandarias, E. Medina and C. Muriel, ``New
symmetry reductions for some ordinary differential equations'',
{\it J. Nonlin. Math. Phys.} {\bf 9} (2002) Suppl.1, 47-58

\bibitem{Olv} P.J. Olver, {\it Application of Lie groups to differential equations}, Springer 1986

\bibitem{PuS} E. Pucci and G. Saccomandi, ``On the reduction methods for ordinary differential equations'', {\it J. Phys. A} {\bf 35} (2002), 6145-6155

\bibitem{Sha} R.W. Sharpe, {\it Differential Geometry}, Springer 1997

\bibitem{Ste} H. Stephani, {\it Differential equations. Their solution using symmetries}, Cambridge University Press 1989

\bibitem{Str} S. Sternberg, {\it Lectures on differential geometry}, Chelsea 1983

\bibitem{Win} P. Winternitz, ``Lie groups and solutions of nonlinear PDEs'', in {\it Integrable systems, quantum groups, and quantum field theory} (NATO ASI 9009), L.A. Ibort and M.A. Rodriguez eds., Kluwer 1993





\end{thebibliography}
\end{document}